# A Pulsar Wind Nebula Embedded in the Kilonova AT 2017gfo Associated with GW170817/GRB 170817A

Jia Ren[1], Da-Bin Lin[1], Lu-Lu Zhang[1], Xiao-Yan Li[1], Tong Liu[2], Rui-Jing Lu[1], Xiang-Gao Wang[1], and En-Wei Liang[1]
[1] Laboratory for Relativistic Astrophysics, School of Physical Science and Technology, Guangxi University, Nanning 530004, People's Republic of China; lindabin@gxu.edu.cn
[2] Department of Astronomy, Xiamen University, Xiamen, Fujian 361005, People's Republic of China

## Abstract

The first detected gravitational wave GW170817 from a binary neutron star merger is associated with an important optical transient AT 2017gfo, which is a direct observation of kilonova. Recent observations suggest that the remnant compact object of the binary neutron star merger associated with GW170817/GRB 170817A may be a stable long-lived magnetized neutron star. In this situation, there would be a pulsar wind nebula (PWN) embedded inside the dynamic ejecta. The PWN emission may be absorbed by the ejecta or leak out of the system. We study the effect of the PWN emission on the observed light curves and radiation spectra. Different from previous works, the absorption and leakage of the PWN emission are all involved in our model, where the absorption of the PWN emission heats up the ejecta and alters its radiation. It is found that the characteristic emission of the embedded PWN quickly evolves. For the multiband and long-term observations of AT 2017gfo, we find that the dynamic ejecta with a PWN emission can fit the observational data very well, especially for the light curves at $t \sim 5$ days and those in the late phase. In addition, our model can naturally generate the thermal to nonthermal spectrum evolution of AT 2017gfo. Our fitting result suggests that a PWN is embedded in the AT 2017gfo.

## 1. Introduction

Compact binary mergers are the main sources of gravitational wave (GW) in the frequency range of the Advanced Laser Interferometer Gravitational-wave Observatory (LIGO) and the Advanced Virgo GW detectors. Among them, the mergers of a binary neutron star (NS) and NS-black hole (BH) draw a lot of attention since they are also potential sources of electromagnetic radiation (EM). GW170817 is the first GW signal from an NS–NS merger detected by the advanced LIGO and Virgo detectors on 2017 August 17 12:41:04 UT (Abbott et al. 2017a). About 2 s after GW170817, the Fermi (Goldstein et al. 2017) and INTEGRAL satellites (Savchenko et al. 2016) detected a short gamma-ray burst (GRB), GRB 170817A, from a location coincident with GW170817. These observations were followed by a detection of an optical counterpart, AT 2017gfo, associated with the accompanying macronova/kilonova powered by the radioactive decay of heavy elements formed in the NS–NS merger (Li & Paczyński 1998; Metzger & Berger 2012; Berger et al. 2013; Fernández & Metzger 2016; Liu et al. 2017; Ma et al. 2018; Song et al. 2018). The accompanying macronova/kilonova was independently confirmed by several teams (e.g., Abbott et al. 2017a; Arcavi et al. 2017; Hu et al. 2017; Lipunov et al. 2017; Smartt et al. 2017; Soares-Santos et al. 2017; Tanvir et al. 2017; Valenti et al. 2017). The observations of AT 2017gfo are performed from a few hours to more than 20 days after the burst trigger until the flux is below the detection threshold. The peak frequency of the spectral energy distribution (SED) in AT 2017gfo is estimated and evolves from UV band to IR band over time. In addition, the SED of AT 2017gfo is found to be thermal-dominated in the early phase and evolves to a nonthermal-dominated one in the late phase (Troja et al. 2017).

The AT 2017gfo is an extremely important source for understanding the physics of kilonova. The idea of kilonova was first introduced by Li & Paczyński (1998). They showed that the radioactive ejecta from an NS–NS or BH–NS merger could power a transient emission and developed a toy model to estimate the light curves. The observations reveal that the early and late phases of AT 2017gfo cannot be consistently explained in the radioactivity-powered kilonova model with a single set of parameters (Cowperthwaite et al. 2017; Kasliwal et al. 2017; Kilpatrick et al. 2017; Shappee et al. 2017; Tanvir et al. 2017; Villar et al. 2017). Therefore, it was widely discussed that the merger ejecta might consist of a multicomponent with different opacity, mass, velocity, and morphology. The AT 2017gfo emission may consist of an early "blue" (light r-process) component and a late "red" (heavy r-process) component (Metzger & Fernández 2014; Kasen et al. 2015, 2017). Summing the light from both a "blue" and "red" component of ejectas provides a comprehensive theoretical model of AT 2017gfo (e.g., Cowperthwaite et al. 2017; Perego et al. 2017; Tanaka et al. 2017; Utsumi et al. 2017; Villar et al. 2017; Kawaguchi et al. 2018; Wanajo 2018; Waxman et al. 2018; Wu et al. 2019). However, the ejecta mass required in the multicomponent model is relatively large, which is hardly meet in an NS–NS merger (Yu et al. 2018). Recent works show that the remnant compact object of GW170817 may be a stable long-lived magnetized NS. In this situation, the magnetic dipole (MD) radiation of the magnetized NS can provide additional energy injected into the ejecta. This process would significantly increase the luminosity of the kilonova (Fan et al. 2013; Yu et al. 2013; Gao et al. 2015). With an additional energy injection from the merger central region, the ejecta mass required to explain the observations can be somewhat smaller than that in the multicomponent model (Li et al. 2018; Yu et al. 2018). In fact, the ejecta with energy injection from a magnetized NS has been studied in explaining the observations of AT 2017gfo (Metzger et al. 2018; Matsumoto et al. 2018; Li et al. 2018; Yu et al. 2018). However, we note that the fitting results of multiband light curves are not quite satisfactory, especially in the late phase of light curves. This issue is presented in most of the published fitting results about AT 2017gfo. Another issue related to AT 2017gfo is the evolution of its SED, which varied from the thermal spectrum in the early phase to a nonthermal spectrum in the late phase. The above issues are our focuses in this work.

We note that the newly formed magnetized NS powers the ejecta via pulsar wind and the interaction between the pulsar wind and the ejecta leading to the formation of the following structures (see Figure 2 of Kotera et al. 2013): a forward shock at the interface between the shocked and unshocked ejectas; a reverse shock at the interface between the shocked and unshocked pulsar wind (commonly called the "termination shock"). The shocked material between the forward and the reverse shocks constitutes the pulsar wind nebula (PWN, e.g., Chevalier & Fransson 1992; Gaensler & Slane 2006; Kotera et al. 2013). The PWN emission has been of interest in the field of GRB (Usov 1992; Blackman & Yi 1998; Dai & Lu 1998; Zhang & Mészáros 2001; Yu & Dai 2007; Lin et al. 2018), supernovae (Vietri & Stella 1998; Inoue et al. 2003), superluminous supernovae (Thompson et al. 2004; Maeda et al. 2007; Kasen & Bildsten 2010; Murase et al. 2015; Wang et al. 2015; Kashiyama et al. 2016), and kilonovae (Gao et al. 2015, 2017; Kisaka et al. 2015; Kisaka et al. 2016). When someone has a view to a PWN-associated kilonova, it is interesting to study how the PWN radiation affects the observed light curve of kilonova, e.g., AT 2017gfo.

Our article is organized as follows. In Section 2 we introduce the model we used. In Section 3 we give the model analysis and apply our model in AT 2017gfo. The multiband and long-term observations are fitted with our model. The conclusions are summarized in Section 4. Throughout this work, we use the notation $Q = 10^x Q_x$ in CGS units unless noted otherwise.

## 2. Dynamics of Ejecta and PWN

Accompanying a mass ejection of a few times $0.01 M_\odot$, either a BH or NS is formed as the final remnant compact object of an NS–NS merger (e.g., Abbott et al. 2017b). The dynamic ejecta is expected to be neutron-rich and thus heavier radioactive elements are synthesized via the rapid neutron capture. The produced heavier elements are unstable and radioactively decay to heat up the ejecta, which results in ultraviolet-optical-infrared emissions. Besides, a newly formed NS deposits extra energy to power the kilonova emission via pulsar winds (Yu et al. 2013; Kasen et al. 2015; Murase et al. 2018). The interaction between the pulsar wind and the ejecta leads to the formation of a PWN between the forward and the reverse shocks (Kotera et al. 2013). The emission of PWN heats up the ejecta or leaks out of the system. Then, the observed flux consists of the emission $F_\nu^b$ from the ejecta and the leaked part $F_\nu^{\rm leak}$ from the PWN, i.e.,

$$F_\nu^{\rm tot} = F_\nu^b + F_\nu^{\rm leak}. \quad (1)$$

The estimations of $F_\nu^b$ and $F_\nu^{\rm leak}$ are given in Sections 2.1 and 2.2, respectively. It should be noted that the energy absorbed from the PWN emission by the ejecta will be reprocessed as thermal energy of the ejecta.

### 2.1. Emission of Ejecta

In this work, the dynamics and emission of the ejecta are implemented based on a simplified radiation transfer model given by Kasen & Bildsten (2010) and Metzger (2017a). The model is described as follows. The merger ejecta is divided into $N (\gg 1)$ layers with different expansion velocity $v_i$, where $v_1 = v_{\min}$ and $v_N = v_{\max}$. The location of the $i$th layer at time $t$ is $R_i = v_i t$ and the mass of $i$th layer is $m_i = \int_{R_i}^{R_{i+1}} 4\pi r^2 \rho_{\rm ej}(r, t) dr$ with (Nagakura et al. 2014)

$$\rho_{\rm ej}(r, t) = \frac{(\delta - 3)M_{\rm ej}}{4\pi R_{\max}^3}\left[\left(\frac{R_{\min}}{R_{\max}}\right)^{3-\delta} - 1\right]^{-1}\left(\frac{r}{R_{\max}}\right)^{-\delta}, \quad (2)$$

where $M_{\rm ej}$ is the total mass of the ejecta and $R_{\max}$ ($R_{\min}$) is the outermost (innermost) radius of the ejecta. The evolution of $R_{\max}$ ($R_{\min}$) is roughly read as $R_{\max} = v_{\max} t$ ($R_{\min} = v_{\min} t$). The ejecta emission is related to the thermal energy $E_i$, of which the evolution can be described as

$$\frac{dE_i}{dt} = (1 - e^{-\Delta\tau_i})e^{[-(\tau_{\rm tot}-\tau_i)]}\xi L_{\rm md}$$
$$+ m_i \dot{q}_{\rm r} \eta_{\rm th} - \frac{E_i}{R_i}\frac{dR_i}{dt} - L_i \quad \text{for } i = 1,\dots, N. \quad (3)$$

Here, the first term on the right-hand side describes the energy absorbed by the $i$th layer for that from the central engine, the second term is the energy released via heavier radioactive element decay in each layer, the third term is the cooling due to the adiabatic expansion, and the last term is the radiation cooling. The detailed information about parameters is presented as follows.

1. The $L_{\rm md}$ is the power of the pulsar wind from the NS and can be estimated by MD radiation, i.e.,

$$L_{\rm md}(t) = L_{\rm md,0}\left(1 + \frac{t}{t_{\rm sd}}\right)^{-\alpha} \quad (4)$$

with

$$L_{\rm md,0} = \frac{B_{\rm p}^2 R^6 \Omega_0^4}{6c^3} = 9.6 \times 10^{42} R_6^6 B_{\rm p,12}^2 P_{0,-3}^{-4} \text{ erg s}^{-1}, \quad (5)$$

where $\Omega_0$, $R$, $B_p$, and $P_0$ are the initial angular frequency, radius, surface polar magnetic field, and initial spin period of the NS, respectively. The spindown timescale $t_{\rm sd}$ and decay index $\alpha$ are related to the energy loss process of NS spindown and can in principle be taken as

$$t_{\rm sd} = \frac{5c^5}{128 GI\epsilon^2\Omega_0^4} = 9.1 \times 10^5 \epsilon_{-4}^{-2} I_{45}^{-1} P_{0,-3}^4 \text{ s with } \alpha = 1 \quad (6)$$

for the GW-dominated spindown loss regime and

$$t_{\rm sd} = \frac{3c^3 I}{B_{\rm p}^2 R^6 \Omega_0^2} = 2.05 \times 10^9 I_{45} B_{\rm p,12}^{-2} P_{0,-3}^2 R_6^{-6} \text{ s with } \alpha = 2 \quad (7)$$

for the MD-dominated spindown loss regime, where $G$ is the gravitational constant, $I$ is the rotational inertia of NS, and $\epsilon$ is the NS ellipticity. In the first term on the right-hand side of Equation (3), $\xi$ describes the fraction of $L_{\rm md}$ which can be absorbed by the ejecta. In addition, $\tau_i$ is the optical depth of the $i$th layer for the observer and can be described as $\tau_i = \sum_i^{N-1}\Delta\tau_i$ with $\Delta\tau_i = \int_{R_i}^{R_{i+1}} \kappa\rho(r) dr$. $\tau_{\rm tot} = \sum_1^{N-1}\Delta\tau_i = \int_{R_{\min}}^{R_{\max}} \kappa\rho(r) dr$ is the total optical depth of the whole ejecta in the line of sight. According to Equation (3), the total energy absorbed by the ejecta is $(1 - e^{-\tau_{\rm tot}})\xi L_{\rm md}$, which is consistent with Equation (21). It should be noted that the emission of the PWN rather than the pulsar spindown luminosity heats up the ejecta.

2. The radioactive power per unit mass $\dot{q}_r$ and the thermalization efficiency of the radioactive power $\eta_{\rm th}$ can be read as (Korobkin et al. 2012; Barnes et al. 2016; Metzger 2017b)

$$\dot{q}_r = 4 \times 10^{18}\left[\frac{1}{2} - \frac{1}{\pi}\arctan\left(\frac{t - t_0}{\sigma}\right)\right]^{1.3} {\rm erg \ s^{-1} \ g^{-1}} \quad (8)$$

and

$$\eta_{\rm th} = 0.36\left[\exp(-0.56 t_{\rm day}) + \frac{\ln(1 + 0.34 t_{\rm day}^{0.74})}{0.34 t_{\rm day}^{0.74}}\right], \quad (9)$$

respectively. Here, $t_0 = 1.3$ s, $\sigma = 0.11$ s, and $t_{\rm day} = t/1$ day.

3. The $L_i$ is the radiation luminosity of the $i$th layer and can be estimated with

$$L_i = \frac{E_i}{\max\{t_d^i, t_{\rm lc}^i\}}, \quad (10)$$

where the photon diffusion timescale $t_d^i$ can be described as

$$t_d^i \simeq \frac{\kappa}{\beta R_i c}\sum_{j=i}^{N-1} m_j \quad (11)$$

and the light-crossing time $t_{\rm lc}^i = R_i/c$. Here, $\beta$ is a numerical factor reflecting the density distribution of the ejecta and $\beta \simeq 13.7$ is adopted in this work (Arnett 1982).[3] For the reason why Equation (10) is a reasonable replacement for a full diffusion discretization, please refer to Huang et al. (2018) for details.

The total bolometric luminosity $L_{\rm bol}$ of the ejecta is estimated by summing the radiation luminosity from all of the layers, i.e.,

$$L_{\rm bol} = \sum_{i=1}^{N-1} L_i. \quad (12)$$

It is always assumed that the radiation of the ejecta is from the photosphere $R_{\rm ph}$ with a blackbody radiation spectrum and the effective temperature $T_{\rm eff}$ is described as (Xiao et al. 2017; Li et al. 2018; Yu et al. 2018)

$$T_{\rm eff} = \left(\frac{L_{\rm bol}}{4\pi\sigma_{\rm SB}R_{\rm ph}^2}\right)^{1/4}, \quad (13)$$

where $\sigma_{\rm SB}$ is the Stephan–Boltzmann constant. The photosphere radius $R_{\rm ph}$ is estimated by setting $\tau_{\rm ph} = \int_{R_{\rm ph}}^{R_{\rm max}}\rho(r)dr = 1$ if $\tau_{\rm tot} > 1$. If $\tau_{\rm tot} \leqslant 1$, we fix $R_{\rm ph}$ to $R_{\rm min}$. The flux density at frequency $\nu$ from the ejecta is given by

$$F_\nu^b = \frac{2\pi h\nu^3}{c^2}\frac{1}{\exp(h\nu/kT_{\rm eff}) - 1}\frac{R_{\rm ph}^2}{D_L^2}, \quad (14)$$

where $h$ is the Planck constant, $k$ is the Boltzmann constant, and $D_L = 40$ Mpc is the luminosity distance of AT 2017gfo.

### 2.2. Emission of the PWN

At the interface between the shocked and unshocked pulsar wind ("termination shock"), electrons and positrons carried in the cold pulsar wind are accelerated to ultrarelativistic energies and the magnetic fields are amplified. The accelerated leptons and the amplified magnetic fields fill the PWN out to the radius $R_{\rm PWN}$. Assuming that the magnetic energy density behind the shock is a fraction $\epsilon_B$ of the total energy density, the magnetic energy density $U_B^{\rm PWN}$ in the PWN can be parameterized as (Tanaka & Takahara 2010, 2013; Murase et al. 2016)

$$U_B^{\rm PWN} = \frac{B_{\rm PWN}^2}{8\pi} = \frac{3}{4\pi}\epsilon_B R_{\rm PWN}^{-3}(t)\int_0^t L_{\rm md}(s)ds. \quad (15)$$

Here, $R_{\rm PWN} \sim R_{\rm min}$ is taken since the deceleration time of the ejecta is $t_{\rm dec} = [3M_{\rm ej}/(4\pi n_{\rm ism}m_p)]^{1/3}/v_{\rm ej} \approx 86$ yr with $M_{\rm ej} \sim 10^{-2} - 10^{-1}M_\odot$, $v_{\rm ej} \sim 0.1 - 0.4c$, the circummerger particle density $n_{\rm ism} \sim 10^{-4}$–$10^{-1}{\rm cm}^{-3}$, and $m_p$ being the proton mass. Following Murase et al. (2015), a broken power law is adopted to describe the energy distribution of injected leptons behind the shock front in the PWN, i.e.,

$$\frac{d\dot{n}_e}{d\gamma_e} \propto \begin{cases}\gamma_e^{-q_1}, & \gamma_m \leqslant \gamma_e < \gamma_b, \\ \gamma_e^{-q_2}, & \gamma_b \leqslant \gamma_e \leqslant \gamma_M,\end{cases} \quad (16)$$

where $q_1 \sim 1 - 2$ ($q_2 \sim 2 - 3$) is the low (high)-energy spectral index, $\gamma_b \sim 10^4 - 10^6$ is the characteristic Lorentz factor of the accelerated leptons in the termination shock, and $\gamma_m$ ($\gamma_M$) is the minimum (maximum) Lorentz factor of leptons.

For the synchrotron emission of the PWN in this work, two break frequencies are relevant. The first break frequency is the characteristic synchrotron frequency corresponding to $\gamma_b$, i.e.,

$$\nu_b \approx \frac{3}{4\pi}\gamma_b^2\frac{q_e B_{\rm PWN}}{m_e c}, \quad (17)$$

where $q_e$ is the charge of leptons. The second break frequency is the synchrotron cooling frequency

$$\nu_c \approx \frac{3}{4\pi}\gamma_c^2\frac{q_e B_{\rm PWN}}{m_e c} \quad (18)$$

with $\gamma_c = 6\pi m_e c/(\sigma_T B_{\rm PWN}^2 t)$ being the cooling Lorentz factor and $\sigma_T$ being the Thomson cross-section (Sari et al. 1998). Owing to this cooling effect, the final energy distribution of leptons may be different from Equation (16). With Equation (16) and considering the relation between $\gamma_c$ and $\gamma_b$, the PWN synchrotron radiation can be described as follows (e.g., Murase et al. 2016). In the fast-cooling regime, i.e., $\nu_c < \nu_b$, the synchrotron emission flux density $L_\nu$ at any

---

[3] One can rewrite Equations (22) and (23) in Arnett (1982) as $t_d^i = [3\kappa/(4\pi\alpha I_M R_i c)]\sum_{j=i}^N m_j$, where the values of $\alpha I_M$ shown in the table 2 of Arnett (1980) are around 3 for different density distribution. Here we chose a typical value for $\alpha I_M$.

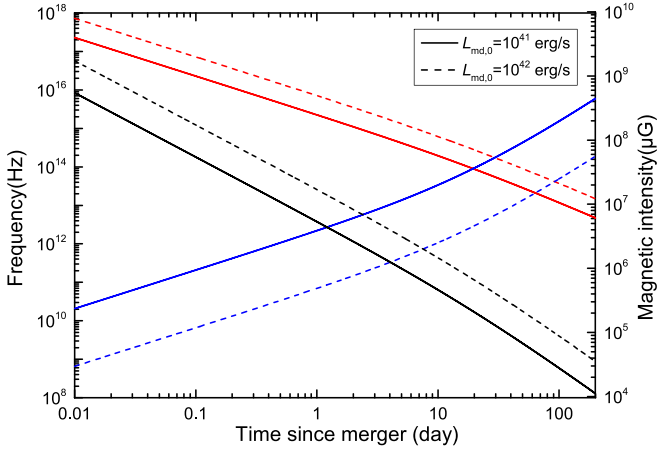

**Figure 1.** Evolutions of $B_{\rm PWN}$ (black line), $\nu_b$ (red line), and $\nu_c$ (blue line) of a PWN associated with a kilonova. Here, $v_{\min} = 0.1c$, $\epsilon_B = 0.01$, $L_{\rm md,0} = 10^{41}$ erg s$^{-1}$ (solid lines), or $10^{42}$ erg s$^{-1}$ (dashed lines), $t_{\rm sd} = 10^6$ s, $\alpha = 1$ and $\gamma_b = 10^4$ are adopted.

frequency $\nu$ can be expressed as

$$\nu L_\nu^{\rm PWN} \approx \frac{\xi L_{\rm md}}{2R_b}$$
$$\times \begin{cases} (\nu_c/\nu_b)^{(2-q_1)/2}(\nu/\nu_c)^{(3-q_1)/2}, & \nu \leqslant \nu_c, \\ (\nu/\nu_b)^{(2-q_1)/2}, & \nu_c \leqslant \nu \leqslant \nu_b, \\ (\nu/\nu_b)^{(2-q_2)/2}, & \nu_b \leqslant \nu \leqslant \nu_M; \end{cases} \quad (19)$$

in the slow-cooling regime, i.e., $\nu_c > \nu_b$, the synchrotron emission flux density can be written as

$$\nu L_\nu^{\rm PWN} \approx \frac{\xi L_{\rm md}}{2R_b}$$
$$\times \begin{cases} (\nu_b/\nu_c)^{(3-q_2)/2}(\nu/\nu_b)^{(3-q_1)/2}, & \nu \leqslant \nu_b, \\ (\nu/\nu_c)^{(3-q_2)/2}, & \nu_b \leqslant \nu \leqslant \nu_c, \\ (\nu/\nu_c)^{(2-q_2)/2}, & \nu_c \leqslant \nu \leqslant \nu_M, \end{cases} \quad (20)$$

where $R_b \sim (2-q_1)^{-1} + (q_2-2)^{-1}$ and the radiation efficiency $\xi = \eta\epsilon_e$ with $\eta = \min\{1, (\nu_b/\nu_c)^{(q_2-2)/2}\}$ (Fan & Piran 2006) is adopted. Based on Equations (19) and (20), one can have $\int_0^{+\infty} L_\nu d\nu \approx \xi L_{\rm md}$. We ignore the effect of the IC scattering process because it is not the main concern of this work. Assuming that the radiation spectrum does not change after photons pass through the ejecta, the observed flux from a PWN can be described as

$$F_\nu^{\rm leak} = \frac{L_\nu e^{-\tau_{\rm tot}}}{4\pi D_L^2}. \quad (21)$$

### 3. Results and Data Fitting

#### 3.1. General Behavior of Kilonova and PWN

The emission of PWN is our main focus in the present work. Then, we first study the evolutions of $B_{\rm PWN}$ (black line), $\nu_b$ (red line), and $\nu_c$ (blue line) for a PWN. The results are shown in Figure 1, where $v_{\min} = 0.1c$, $\epsilon_B = 0.01$, $L_{\rm md,0} = 10^{41}$ erg s$^{-1}$ (solid lines), or $10^{42}$ erg s$^{-1}$ (dashed lines), $t_{\rm sd} = 10^6$ s, $\alpha = 1$, and $\gamma_b = 10^4$ are adopted. One can find that the magnetic field strength decreases rapidly as time goes on. It is very different from that in the PWN associated with a core-collapse supernova (e.g., Figure 4 of Kotera et al. 2013). Then, one would find a PWN with quickly evolving behavior in its radiation. Figure 1 shows that the characteristic synchrotron frequency $\nu_b$ quickly decreases from X-ray band ($t \sim 0.01$ day) to ultraviolet-optical-infrared bands ($t \sim 0.1-10$ days) and even to radio band for the late phase. However, the cooling frequency $\nu_c$ quickly increases from radio band to optical band. Thus, the radiation efficiency of PWN may be low in the late phase. In other words, the radiation of PWN in the late phase would be in the slow-cooling regime. Figure 1 also reveals that a higher energy injection from the central NS would produce a PWN with higher $B_{\rm PWN}$ and $\nu_b$ but lower $\nu_c$. Then, the transition from the fast-cooling regime to the slow-cooling regime would be deferred for a PWN with high energy injection. Since the characteristic synchrotron frequency enters into the optical bands quickly, Equation (3) can provide a good description about the evolution of the internal energy for ejecta.[4] This is different from the situation in which the PWN characteristic synchrotron frequency is in the X-ray band.

In the following two paragraphs, we study the effect of PWN emission on the observed light curves and radiation spectrum. The properties of the ejecta are described with $M_{\rm ej} = 0.03 M_\odot$, $\kappa = 5$ g cm$^{-2}$, $v_{\max} = 0.3c$, and $\delta = 2$. We typically chose $q_1 = 1.8$ and $q_2 = 2.2$ for the PWN. Figure 2 plots the observed flux in the situations with $L_{\rm md,0} = 0$ (black lines), $10^{41}$ erg s$^{-1}$ (red lines), $10^{42}$ erg s$^{-1}$ (orange lines), and $10^{43}$ erg s$^{-1}$ (blue lines). The upper panels show the observed light curves at $K$ (left panel), $F606W$ (middle panel), and $U$ bands (right panel), respectively. Here, the monochromatic AB magnitude is estimated with $M_\nu = -2.5\log_{10}(F_\nu/3631{\rm Jy})$. In these panels, the solid lines plot the behavior of $F_\nu^{\rm tot}$ and the dashed lines depict the evolution of $F_\nu^b$. Comparing the situations with $L_{\rm md,0} = 0$ and $L_{\rm md,0} \neq 0$, one can find that the ejecta emission are very different. The energy injection from the central region can elevate the observed intensity effectively. The higher the value of $L_{\rm md,0}$ adopted, the higher the ejecta luminosity would be. More importantly, the effect of energy injection on the light curves is different for different observational bands. The observed flux smoothly increases in the infrared $K$-band, but a bimodal light curve is presented in the light curves of ultraviolet $U$-band or optical $F606W$-band. The bimodal light curves have also been found in Metzger (2017b). The bimodal light curves become obvious if the injection energy is significantly high. Then, one can believe that the appearance of bimodal structure in the light curves is related to the heating of ejecta by the PWN. In the early phase, the optical depth is high, the PWN emission is mainly absorbed by the inner region of the ejecta and the radiative cooling of the ejecta mainly appears in the outer part. That is to say, the energy injection in the early phase is mainly deposited in the inner layers. With the decrease of optical depth $\tau_{\rm tot}$ (see Figure 3), the deposited energy would be released. Then, one can find a bump in the light curves at the time of $\tau_{\rm tot} \sim 1$. This is the reason for the bimodal light curve behavior. It is interesting to point out that the deposited energy is associated with the value of $L_{\rm md,0}$. Then, the higher $L_{\rm md,0}$ adopted, the more luminosity of the bump at $\tau_{\rm tot} \sim 1$ would be. This is consistent with the light curves shown in Figure 2.

---
[4] Due to the bound–free or bound–bound absorption of the ejecta, the optical depth of X-ray photons are much higher than that of optical photons. Since $\nu_b$ falls in the X-ray bands in the very early phase, the absorption of the PWN emission is more intense than that presented in Equation (3).

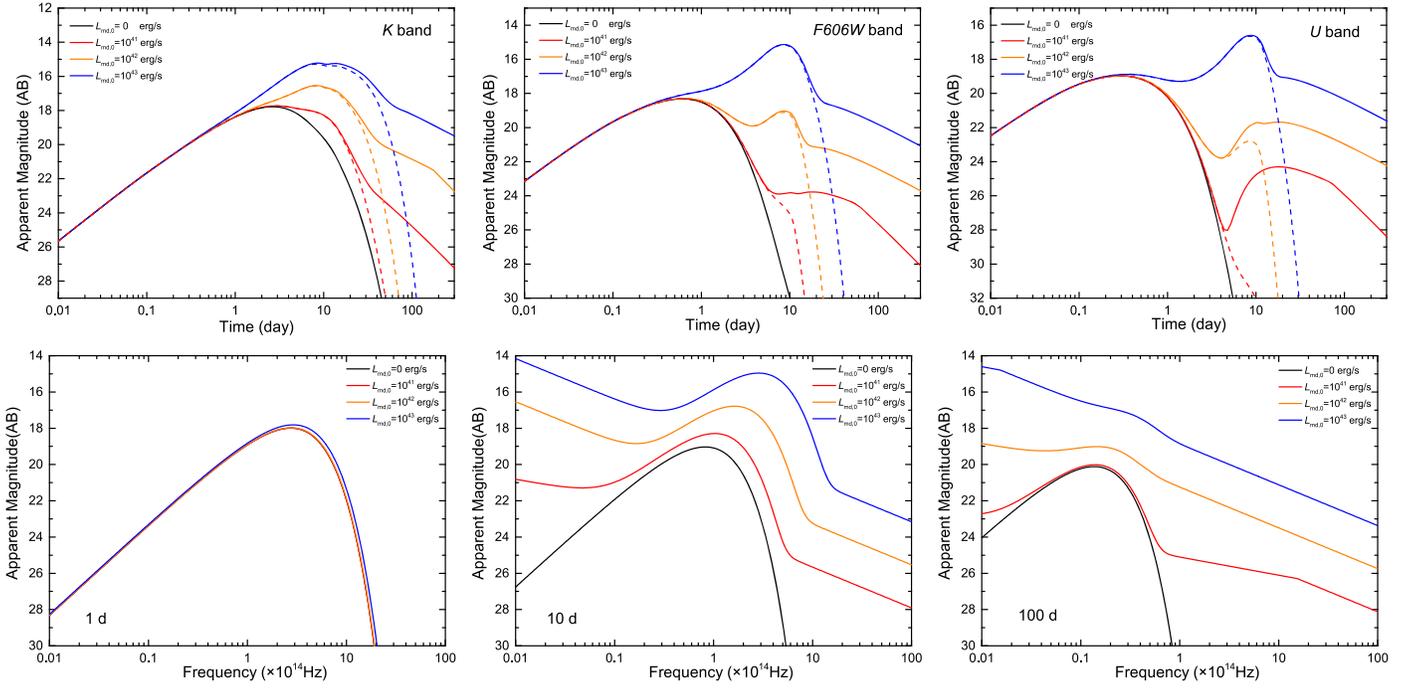

**Figure 2.** Light curves (upper part) and radiation spectra (bottom part) under the situations with different energy injection, i.e., $L_{md,0} = 0$ (black lines), $10^{41}$ erg s$^{-1}$ (red lines), $10^{42}$ erg s$^{-1}$ (orange lines), and $10^{43}$ erg s$^{-1}$ (blue lines). The light curves of $K$-band, $F606W$-band, and $U$-band are shown in the left, middle, and right panels of the upper part, where the dashed lines plot the light curves of thermal emission from the ejecta. The radiation spectra at time $t = 1$ day, 10 days, and 100 days are shown in the left, middle, and right panels of the bottom part, respectively.

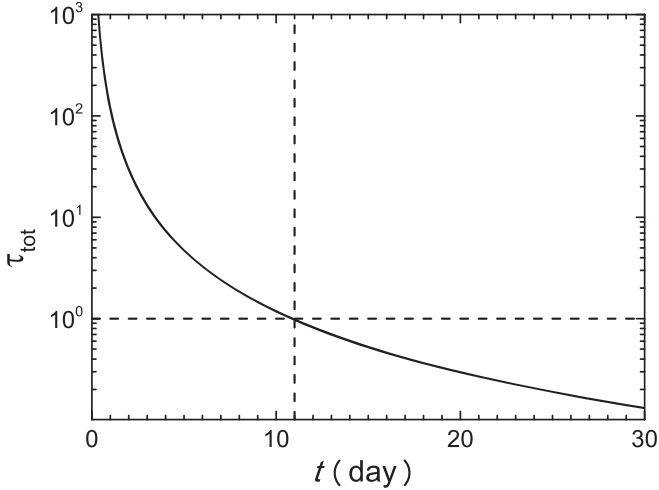

**Figure 3.** Evolution of $\tau_{\rm tot}$ for the ejecta adopted in Figure 2, where the horizontal dashed line indicates $\tau_{\rm tot} = 1$ and the vertical dashed line indicates the corresponding observed time. It can be found that the second peak of the light curves in Figure 2 appears at around the time of $\tau_{\rm tot} = 1$.

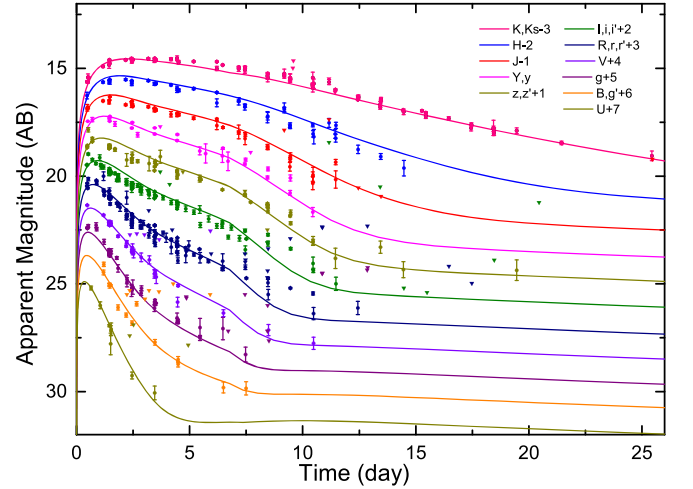

**Figure 4.** Fitting result (solid lines) of AT 2017gfo, where the observational data are described with circles and the upper limits are represented with triangles.

The lower panels of Figure 2 plot the observed radiation spectrum at $t = 1$ day (left panel), 10 days (middle panel), and 100 days (right panel), respectively. As one can find from these panels, the radiation spectrum is the superposition of a thermal component and a nonthermal (power-law) component. Over time the spectral peak of the blackbody component moves toward the lower frequency and its intensity gradually decreases. The nonthermal component emerges at a certain time and becomes dominant in the late phase. These behaviors can be understood as follows. Since the optical depth $\tau_{\rm tot}$ is significantly high in the very early phase, the emission of the PWN is mainly absorbed by the inner layers of the ejecta. In addition, the photon diffusion timescale $t_d$ of inner layers is sufficiently large. Then, the observed radiation is mainly from the outer layers of the ejecta and the effect of the PWN emission cannot be observed. Correspondingly, the radiation spectrum would be a thermal form. This is the reason for the same behavior in the very early phase ($t \lesssim 1$ days) presented in the situations with different $L_{\rm md,0}$. As time goes on, the optical depth gradually decreases and thus the effect of PWN emission begins to emerge. However, the optical depth $\tau_{\rm tot}$ ($\gtrsim 1$) is still high in this phase. In addition, the deposited energy in the inner region of the ejecta is significantly large. Then, the observed flux may be also dominated by the thermal emission of the ejecta rather than the emission from the PWN. This is consistent with those shown in the lower panels of Figure 2 at $t = 10$ days. One can also find this behavior by comparing the light curves of $F_\nu^{\rm tot}$ and $F_\nu^{\rm b}$ in the upper panels of Figure 2. After that, the optical depth decreases to less than unity and thus the thermal energy of the ejecta is almost all released. Then, the observed radiation would be mainly from the PWN. Correspondingly, the radiation spectrum would be dominated by the nonthermal power-law component. It can be found in the radiation spectrum of Figure 2 at $t = 100$ days.

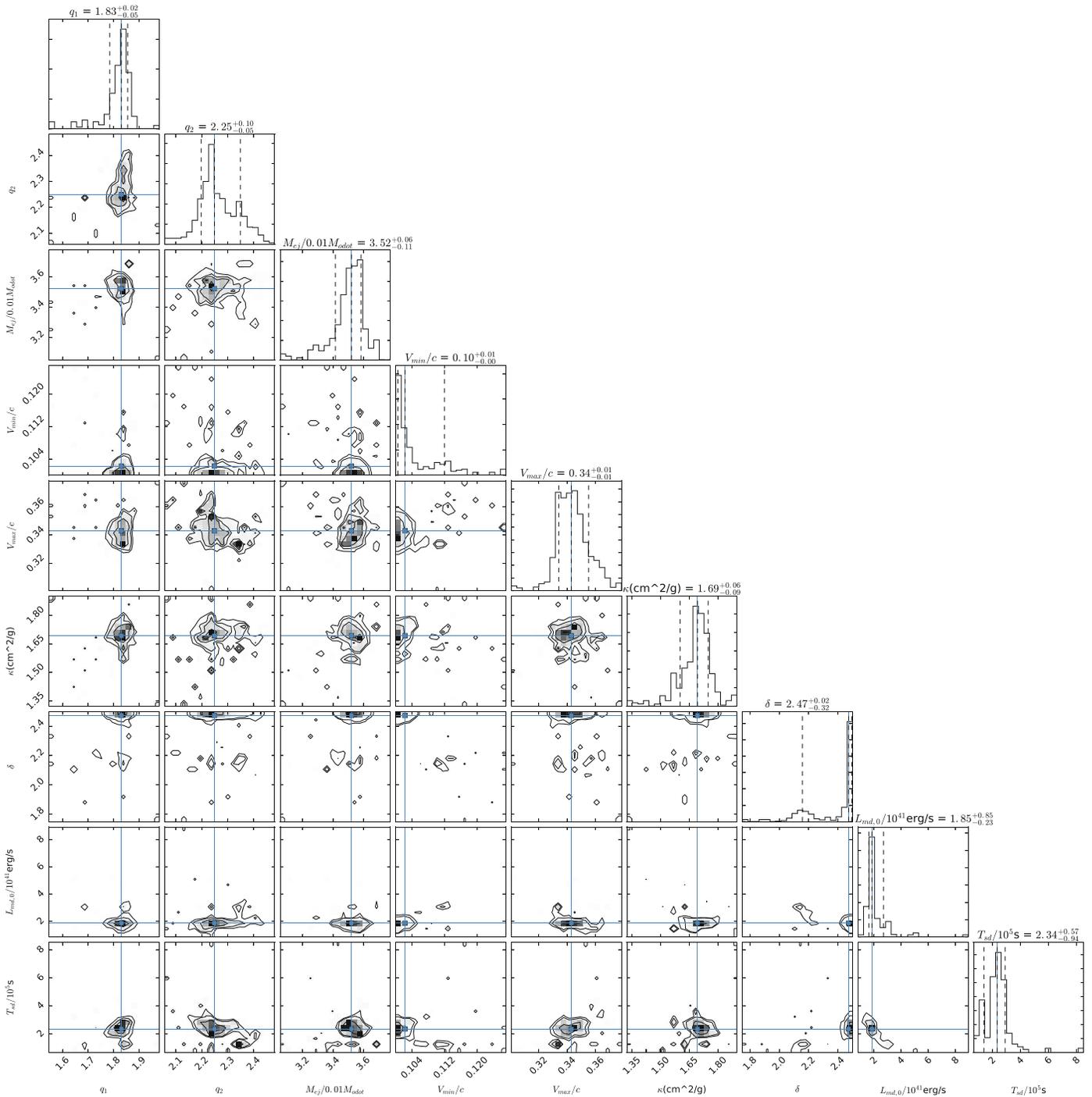

**Figure 5.** Posterior probability density functions for the physical parameters of AT 2017gfo from MCMC simulations.

### 3.2. Application on AT 2017gfo

In this section, we use our model to fit the multiband observations of AT 2017gfo. The observational data of AT 2017gfo are taken from Villar et al. (2017), where the data from *Hubble Space Telescope* (*HST*), *Swift*, and *MASTER* optical detection are not considered in our fittings. The data used in our fits can be found in Figure 5. Our fitting is performed based on the Markov Chain Monte Carlo (MCMC) method to produce posterior predictions for the model parameters. Since 11 light curves on different bands should be fitted simultaneously, we adopt a functional form of the log-likelihood by considering the time averaged. The procedure is shown as follows.

**Table 1**
Fitting Result from MCMC Fitting

| Parameter | Constraint |
|---|---|
| $L_{\rm md,0}/10^{41}$ erg s$^{-1}$ | $1.85^{+0.85}_{-0.23}$ |
| $t_{\rm sd}/10^5$ s | $2.34^{+0.57}_{-0.94}$ |
| $M_{\rm ej}/0.01 M_\odot$ | $3.52^{+0.06}_{-0.11}$ |
| $\kappa/$g cm$^{-2}$ | $1.69^{+0.06}_{-0.09}$ |
| $v_{\rm min}/c$ | $0.10^{+0.01}_{-0.00}$ |
| $v_{\rm max}/c$ | $0.34^{+0.01}_{-0.01}$ |
| $\delta$ | $2.47^{+0.02}_{-0.32}$ |
| $q_1$ | $1.83^{+0.02}_{-0.05}$ |
| $q_2$ | $2.25^{+0.10}_{-0.05}$ |
| $\alpha$ | 1 |

We first divide the observational time into a time serial [0, 1d), [1d, 2d), ⋯, [j − 1d, jd), ⋯, [25d, 26d]. For each observational band (e.g., the ith band), we calculate the number of data points $N_{i,j}$ in the time interval [j − 1d, jd). Then, the log-likelihood functional form is

$$\ln \mathcal{L} = -\frac{1}{2}\sum_{i,j}\left[\frac{1}{M_i N_{i,j}} \times \frac{(O_{i,j} - O^{\rm mod}_{i,j})^2}{\sigma^2_{i,j}} + \ln(2\pi\sigma^2_{i,j})\right], \quad (22)$$

where $M_i$ represents the accumulated number of time intervals with observational data points for ith band and $O_{i,j}$, $O^{\rm mod}_{i,j}$, and $\sigma_{i,j}$ are the observed magnitudes, model magnitudes, and observed uncertainties, respectively. For the upper limit data points, e.g., the last two data points of J-band, if the theoretical results are above these data points, we multiply the obtained $\ln \mathcal{L}$ by 100 as the final returned value. In addition, the observational data points in the high frequency bands are scarce, e.g., B-band and U-band. Then, the last two observational data points of B-band are used to constrain the model. That is to say, if the theoretical results are not located in the error bars of these two data points, the obtained $\ln \mathcal{L}$ is also multiplied by 100 as the final. We use the emcee[5] code for our MCMC fits (Foreman-Mackey et al. 2013), where $n_{\rm walkers} \times n_{\rm steps} = 100 \times 600$ is adopted and 10% of the chain in the start are eliminated. In addition, $\epsilon_B = 0.01$, $\epsilon_e = 1 - \epsilon_B$, and $\gamma_b = 10^4$ are set in our fits.

The posterior probability density functions for the physical parameters of AT 2017gfo from MCMC fits are presented in Figure 4. The optimal result from MCMC fits is shown in Figure 5 with solid lines and the obtained parameters at the 1σ confidence level are reported in Table 1. Figure 5 reveals very good fits of the multiband observations from early phase to late phase. The obtained properties of the ejecta reported in Table 1, i.e., $M_{\rm ej} = 3.52^{+0.06}_{-0.11} \times 10^{-2} M_\odot$, $\kappa = 1.69^{+0.06}_{-0.09}$ cm$^2$ g$^{-1}$, and $v_{\rm min}$ or $v_{\rm max} \sim (0.1$–$0.34)c$, are consistent with those found in previous works (Arcavi et al. 2017; Cowperthwaite et al. 2017; Kasliwal et al. 2017; Kilpatrick et al. 2017; Smartt et al. 2017; Villar et al. 2017). For the value of $\alpha$, $\alpha = 1$, or $\alpha = 2$ are taken in our MCMC fits. However, the MCMC procedure can provide a good fitting of the observational data only in the situation with $\alpha = 1$. It suggests that the spindown energy loss of the remnant NS after GW170817 is in the GW-dominated spindown loss regime, which is consistent with that found in Piro et al. (2019). With the obtained $L_{\rm md,0}$ and $t_{\rm sd}$, the surface magnetic field $B_p = 1.39 \times 10^{11} R_6^{-3} P_{0,-3}^2$ G and the ellipticity $\epsilon = 1.97 \times 10^{-4} I_{45}^{-1/2} P_{0,-3}^2$ of the remnant NS are estimated and consistent with those found in Yu et al. (2018), Piro et al. (2019), and Ai et al. (2018). We note that the obtained optimal pulsar luminosity $L_{\rm md,0} = 1.85^{+0.85}_{-0.23} \times 10^{41}$ erg s$^{-1}$ is low compared to previous studies, e.g., Li et al. (2018). This is due to the higher ejecta mass we have obtained compared with previous studies. Based on the discussion in Section 3.1, the properties (e.g., mass) of the ejecta are mainly associated with the observational data near the peak time (∼1 day) and the optimal pulsar luminosity mainly depends on the late observations (i.e., a few days after the peak time). If a high ejecta mass is adopted in the fittings, the contribution of the PWN emission to the late phase may be low. Then, a low pulsar luminosity may be found if a high ejecta mass is adopted in the fittings.

---
[5] The algorithm of emcee is based on Goodman & Weare's Affine Invariant Markov chain Monte Carlo Ensemble sampler (Goodman & Weare 2010). The reader can refer to Foreman-Mackey et al. (2013) and https://emcee.readthedocs.io/en/stable/ for details.

In order to test our fits result, we confront our result with *HST* observations (left panel), *Swift* observations (middle panel, *UVW1*, *UVW2*, and *UVM2* bands), *MASTER* optical observations (middle panel, *W* band), and *Spitzer Space Telescope* observations (right panel) in Figure 6. We add the observational data of *HST* at >100 days (Lyman et al. 2018; Margutti et al. 2018; Lamb et al. 2019; Piro et al. 2019). It is found that the model curve is lower than data points, which suggests the late time observation is dominated by the afterglow of GRB 170717A. The observational data of the *Spitzer Space Telescope* is from Villar et al. (2018). This figure reveals that our result is consistent with the data from *HST*, *Swift*, *MASTER*, and *Spitzer Space Telescope*, which were not used in our MCMC fits. In observations, it is found that the observed radiation spectrum of AT 2017gfo gradually deviates from the thermal radiation spectrum over time (Troja et al. 2017). Then, we plot the radiation spectrum from 11.5 to 19.5 days after trigger based on our fits result. The radiation spectra are shown in the left panel of Figure 7, where the circles and triangles are the observational data laying closest to our selected time. One can find that our fits result describes the observed radiation spectra nicely. It is interesting to point out that the spectral peak of the thermal component moves from $\sim 10^{14}$ Hz at $t = 10.5$ days to the very low frequency at $t = 19.5$ days. Moreover, the thermal component is comparable to the nonthermal component at $t \lesssim 19.5$ days. This behavior suggests that the bolometric luminosity at $t \lesssim 19.5$ days can be well described with the thermal emission from the ejecta. In the right panel of Figure 7, we compare the thermal luminosity based on our fitting result with the observed bolometric luminosity (Waxman et al. 2018). One can find that our thermal luminosity describes the observed bolometric luminosity very well.

## 4. Summary and Conclusions

Recent works suggest that the remnant compact object of the NS–NS merger producing GW170817/GRB 170817A may be a long-lived magnetized NS (Lü et al. 2019; Piro et al. 2019). In this situation, a pulsar wind is generated by the central magnetized NS and the interaction between the pulsar wind and the ejecta can form a rapidly evolving PWN. The radiation of PWN heats up the ejecta or leaks out of the system and thus can affect the observed light curves. In this work, we study the effect of PWN on the observed light curves and radiation spectrum. It is found that the radiation spectrum evolves from a thermal-dominated radiation spectrum in the early phase to a nonthermal-dominated radiation spectrum at several days. This is consistent with those found in AT 2017gfo. Then, we perform MCMC fits to the multiband observations of AT 2017gfo. It is shown that our model presents a very good description about the multiband light curves of AT 2017gfo. The obtained properties of the ejecta and the magnetized NS are consistent with previous works.

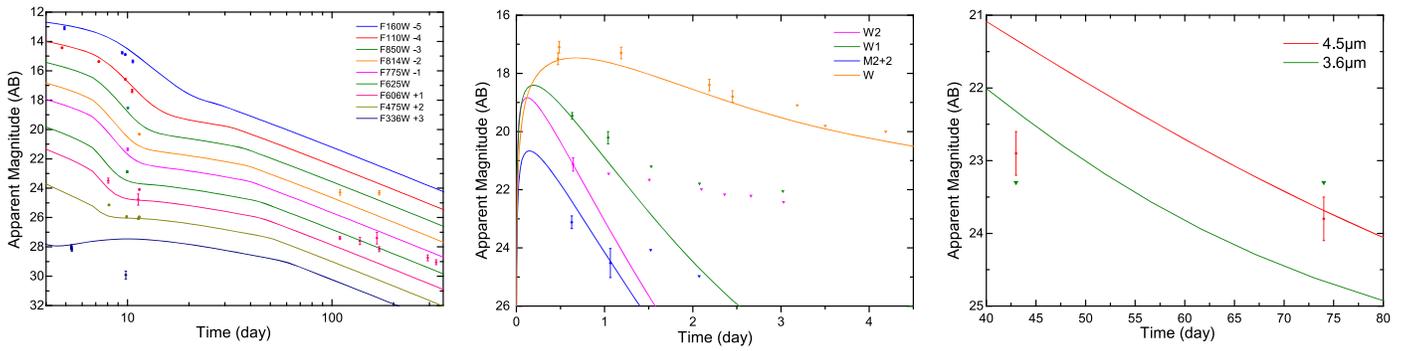

**Figure 6.** Confronting our fitting result with *HST* observations (left panel), *Swift* observations (middle panel, *UVW1*, *UVW2*, and *UVM2* bands), *MASTER* optical observations (middle panel, *W* band), and *Spitzer Space Telescope* observations (right panel).

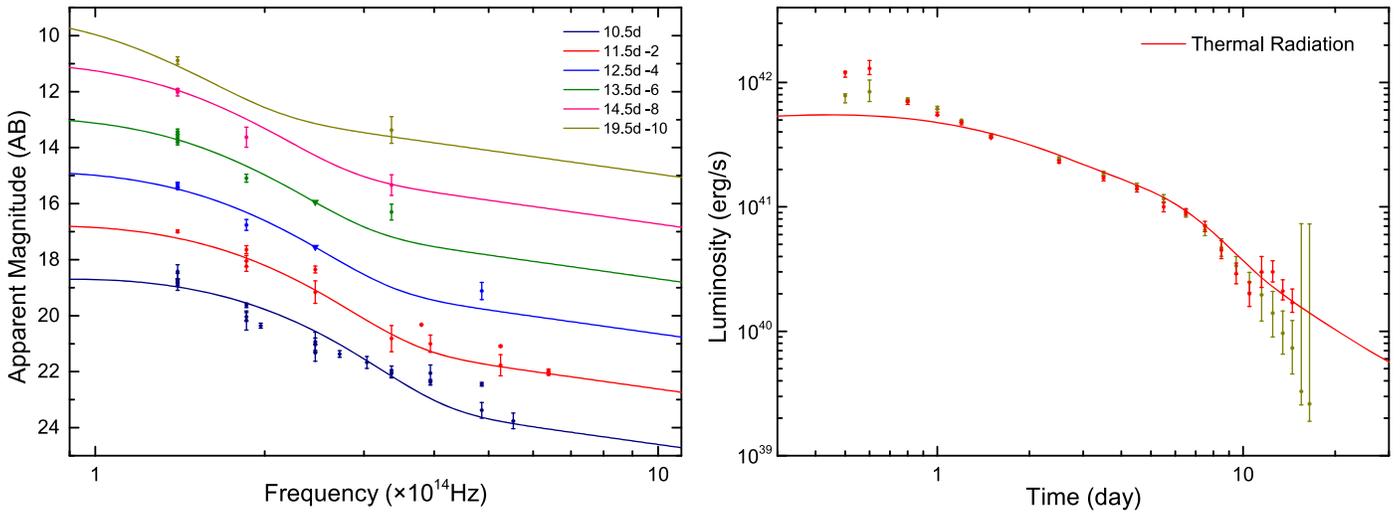

**Figure 7.** Comparison of radiation spectra (left panel) and bolometric luminosity (right panel).

We thank the anonymous referee of this work for beneficial suggestions that improved the paper. We also thank Shan-Qin Wang and Yuan-Pei Yang for helpful discussions and suggestions. This work is supported by the National Natural Science Foundation of China (grant Nos. 11773007, 11533003, 11822304, 11851304, 11673006, U1731239, U1938201), the Guangxi Science Foundation (grant Nos. 2018GXNSFFA281010, 2016GXNSFDA380027, 2017AD22006, 2018GXNSFGA281005, 2018GXNSFDA281033, 2016GXNSFFA380006), and the Innovation Team and Outstanding Scholar Program in Guangxi Colleges and the Innovation Project of Guangxi Graduate Education (grant No. YCSW20 18050).

*Software:* emcee Foreman-Mackey et al. (2013), matpltlib Hunter (2007), numpy Van Der Walt et al. (2011).

### ORCID iDs

Jia Ren ● https://orcid.org/0000-0002-9037-8642
Da-Bin Lin ● https://orcid.org/0000-0003-1474-293X
Tong Liu ● https://orcid.org/0000-0001-8678-6291
Xiang-Gao Wang ● https://orcid.org/0000-0001-8411-8011
En-Wei Liang ● https://orcid.org/0000-0002-7044-733X